# Single-layer MoS$_2$ roughness and sliding friction quenching by interaction with atomically flat substrates

J. Quereda,[1] A. Castellanos-Gomez,[2] N. Agraït,[1,3,4,5] and G. Rubio-Bollinger[1,4,5,a]

[1] *Departamento de Física de la Materia Condensada. Universidad Autónoma de Madrid, Madrid, E-28049, Spain.*

[2] *Kavli Institute of Nanoscience, Delft University of Technology, Lorentzweg 1, 2628 CJ Delft (The Netherlands).*

[3] *Instituto Madrileño de Estudios Avanzados en Nanociencia, IMDEA-Nanociencia, E-28049, Madrid, Spain.*

[4] *Instituto de Ciencia de Materiales Nicolás Cabrera, Campus de Cantoblanco, E-28049, Madrid, Spain.*

[5] *Condensed Matter Physics Center (IFIMAC), Universidad Autónoma de Madrid, E-28049 Madrid, Spain*

We experimentally study the surface roughness and the lateral friction force in single-layer MoS$_2$ crystals deposited on different substrates: SiO$_2$, mica and hexagonal boron nitride (h-BN). Roughness and sliding friction measurements are performed by atomic force microscopy (AFM). We find a strong dependence of the MoS$_2$ roughness on the underlying substrate material, being h-BN the substrate which better preserves the flatness of the MoS$_2$ crystal. The lateral friction also lowers as the roughness decreases, and attains its lowest value for MoS$_2$ flakes on h-BN substrates. However, it is still higher than for the surface of a bulk MoS$_2$ crystal, which we attribute to the deformation of the flake due to competing tip-to-flake and flake-to-substrate interactions.

In the last few years, a wide family of novel two-dimensional crystals has been investigated [1], showing a large variety of different electrical and mechanical behaviors [2]. Atomically thin MoS$_2$ [3,4-6], for instance, has been proposed as an attractive two-dimensional material due to its large intrinsic bandgap of 1.8 eV, well-suited for electronics and optoelectronics applications [4,7]. Further, atomically thin MoS$_2$ crystal properties such as photoluminescence [8], electrostatic screening [9] or mechanical behavior [5,10] have been recently studied.

Due to their inherent large surface-to-volume ratio, the chemical, optical and electrical properties of atomically thin materials can be strongly modified by their interaction with the substrate where they are deposited [11,12]. For instance, it is well-known that graphene crystals deposited on standard SiO$_2$ substrates exhibit different characteristics from those observed for free-standing crystals due to increased topographic corrugation and spatial charge inhomogeneities present at the graphene/substrate interface [13]. As a consequence, graphene devices fabricated on SiO$_2$ substrates present an electron mobility below the theoretical prediction for a free-standing geometry [14]. This problem can be overcome to a large extent by employing atomically flat substrates with low density of charged impurities [15-17]. For example, hexagonal boron nitride (h-BN) has demonstrated to be an excellent substrate to minimize the corrugation and electronic inhomogeneities in graphene devices [18,19]. In fact, mobility measurements in graphene devices on h-BN have reached values of $60,000$ cm$^2$V$^{-1}$s$^{-1}$ [18].

---

[a] Author to whom correspondence should be addressed. Electronic mail: gabino.rubio@uam.es

Here we compare the surface roughness, measured by atomic force microscopy (AFM), of single-layer MoS$_2$ crystals deposited either on amorphous or on atomically flat crystalline substrates. We find that the single-layer MoS$_2$ follows quasi-conformally the topography of the substrate and thus its roughness is dominated by the roughness of the substrate. Therefore, amorphous SiO$_2$ substrates induce a much larger corrugation in the MoS$_2$ monolayers than atomically flat substrates. In fact, MoS$_2$ monolayers transferred onto mica or h-BN substrates present a roughness about 50% lower than on SiO$_2$. Nonetheless, while the MoS$_2$ monolayer on h-BN shows a roughness comparable to that of pristine bulk MoS$_2$, single-layer MoS$_2$ on mica presents a noticeable larger roughness. We attribute this larger roughness to the presence of potassium carbonate crystallites [20] and water molecules trapped at the MoS$_2$/mica interface due to the highly hydrophilic character of mica. We find a marked correlation between the substrate-induced roughness and the friction in MoS$_2$ layers, indicating that the friction in atomically thin MoS$_2$ layers can be strongly modified by the MoS$_2$/substrate interaction.

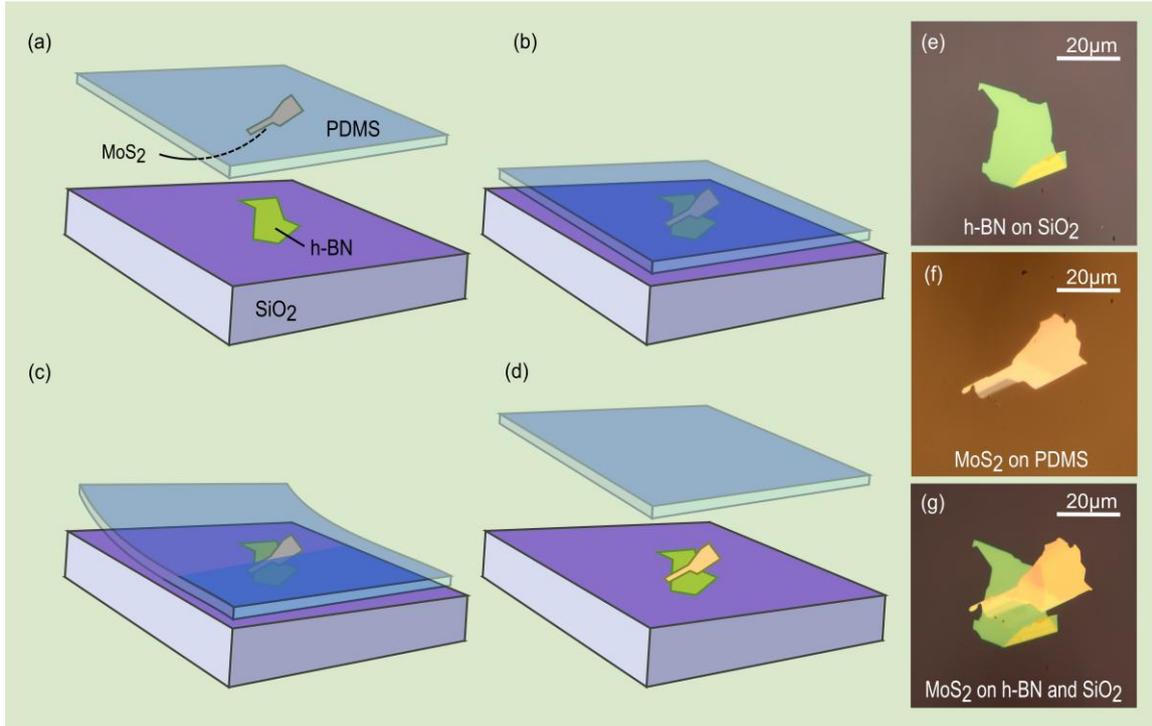

FIG. 1. Schematic cartoon (a-d) and optical micrographs (e-g) of the crystal transfer process with viscoelastic stamps. After preparing a h-BN flake on the $SiO_2$ surface (e), an atomically thin $MoS_2$ crystal on the stamp (f) with a one layer thick region is identified. Then, the $MoS_2$ flake is aligned with the h-BN crystal (a) and the stamp is gently lowered to put in contact both flakes (b). If the stamp is then very slowly separated from the substrate (c), the $MoS_2$ thin flake is left on top of the h-BN crystal (d). (g) Optical micrograph of the $MoS_2$ flake deposited on the h-BN flake.

Single-layer $MoS_2$ flakes are deposited onto a $SiO_2$/Si substrate (amorphous) and onto h-BN and mica substrates (crystalline and atomically flat). For the case of the $SiO_2$/Si substrate, the flakes are deposited by mechanical exfoliation using viscoelastic poly-dimethyl siloxane stamps. The use of viscoelastic stamps instead of the standard adhesive Scotch or Nitto tape provides a cleaner all-dry method for the flake transfer, avoiding the presence of adhesive traces in the resulting sample [6,16,21]. For the case of the h-BN or mica substrates, h-BN or mica flakes are firstly deposited onto a $SiO_2$/Si substrate and then we use a deterministic all-dry transfer method, shown in Figures 1a – 1d, to transfer the $MoS_2$ flakes on top of the h-BN or mica flakes [21]. The h-BN or mica flakes are inspected prior the transfer by optical microscopy to select flat and wrinkle-free flakes (Figure 1e). Moreover, h-BN or mica flakes with a thickness between 20 nm to 60 nm are selected to ensure that the flakes are stiff enough to lay flat of the surface without following the corrugation of the underlying $SiO_2$/Si substrate and to decouple from charged impurities in the $SiO_2$ substrate. The second step is to deposit a single layer $MoS_2$ flake on top of the selected mica or h-BN flake. We follow the procedure described in previous works [16,21] to achieve a deterministic and controlled transfer of the $MoS_2$ monolayer. The procedure relies on the convenient optical and mechanical properties of the viscoelastic stamp used for the transfer. First, we exfoliate repeatedly a MoS$_2$ crystal using Nitto tape. Once we reach sufficiently thin crystals (<10 nm) we exfoliate and pick them up from the Nitto tape using a viscoelastic stamp, obtaining some atomically thin MoS$_2$ flakes attached to the stamp, as shown in Figure 1f. Because the stamp is transparent, these flakes can be inspected under an optical microscope using white light transmission illumination. MoS$_2$ single layers can be identified on the stamp by their characteristic small but detectable optical contrast (C = 0.06±0.01) [22]. The MoS$_2$ flake under the stamp is then aligned with the selected mica or h-BN crystal using a three axis XYZ micromanipulator while being observed under the microscope. Then, the stamp is lowered, gently putting into contact the MoS$_2$ flake and the mica or h-BN flake. Subsequently, the stamp is slowly lifted and detached from the substrate, leaving the MoS$_2$ flake lying on top of the mica or h-BN flake (Figure 1g).

The thickness of the transferred flakes has been determined by contact mode atomic force microscopy under ambient conditions using a cantilever with a flexural elastic constant of 0.76 N/m (Olympus OMCL-RC800PSA). Contact mode AFM enables to simultaneously measure surface roughness and sliding friction and prevents from topography artifacts [23]. Figure 2 shows the topography and friction images acquired at the MoS$_2$/substrate edges for SiO$_2$, mica and h-BN substrates. In Figure 2a we show the topography of a single-layer of MoS$_2$ on SiO$_2$ and in Figure 2g a single line scan. We quantitatively obtain the roughness of the MoS$_2$ surface from height histograms of topography images, and analyzing the standard deviation (σ) of a fit to a Gaussian distribution, as shown in Figure 3. We find that the roughness of the SiO$_2$ substrate is σ = 190 pm, while the roughness of a single-layer MoS$_2$ flake on SiO$_2$ (σ = 108 pm) is significantly lower due to the bending rigidity and high elastic modulus of MoS$_2$ [5], but still much larger than the roughness of a pristine surface of a bulk MoS$_2$ crystal (σ = 66 pm). The roughness ratio between single-layer MoS$_2$ on SiO$_2$ and bare SiO$_2$ is 0.56 while the equivalent ratio reported for graphene is 0.9 [15]. The difference can be attributed to the dissimilar bending rigidities of single-layer MoS$_2$ [24] and graphene [25] which are 9.6 eV and 1.4 eV respectively.

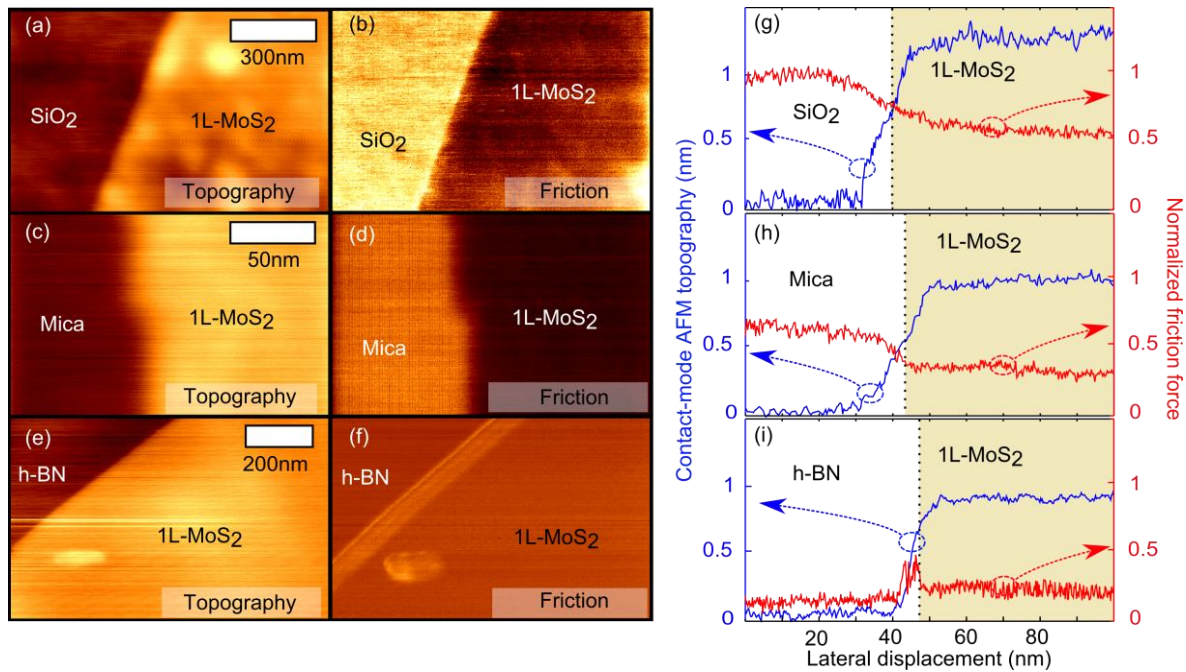

FIG. 2. Contact-mode AFM topography and lateral friction force images of a single layer MoS$_2$ crystal (1L-MoS$_2$) on SiO$_2$, (a, b), mica (e, f) and h-BN (c, d). The friction of the MoS$_2$ layer region is lower than that of the substrate for the case of SiO$_2$ (b) and mica (f), while for the case of the h-BN substrate the friction on the MoS$_2$ flake is slightly higher (d). (g - i) AFM topography (blue, left axis) and friction force (red, right axis) profiles, as a function of tip lateral displacement, across regions comprising both the bare substrate of SiO$_2$ (g), mica (h) or h-BN (i) and a single-layer MoS$_2$ flake laying on the substrate. The surface roughness differences can be readily observed for the MoS$_2$ on the various substrates and is quantitatively shown in Figures 3 and 4. Differences in friction are also apparent and their values are shown in Figure 4. Units of friction force are normalized to their mean value for the bare SiO$_2$ substrate, which in this case is 260 nN.

We further study the roughness of MoS$_2$ monolayers deposited on mica substrates (Figures 2e and 2h). Although mica is an atomically flat crystal we measure a surface roughness of σ = 94 pm for the bare mica surface. This unexpectedly large roughness for a perfectly cleavable surface can be attributed to the presence of water and the formation of nano-crystallites, like potassium carbonate, on the mica surface when cleaved in air [20]. In fact, a much lower roughness for both graphene flakes on mica and bare mica surfaces has been reported when using a controlled atmosphere, free of oxygen and water, to prevent deterioration of the mica surface [15,26]. The roughness of a single layer of $MoS_2$ on mica, $\sigma = 77$ pm, is much smaller than on $SiO_2$ and slightly lower than for bare mica, but still not as low as for bulk $MoS_2$. Therefore, as for the $SiO_2$ substrate, the roughness of the $MoS_2$ sheet significantly follows the topography of the mica substrate.

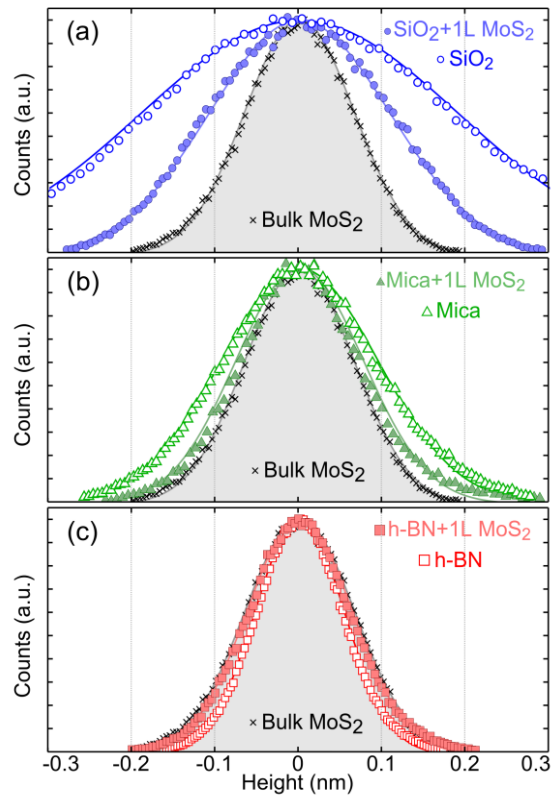

FIG. 3. Height histograms of 100 nm x 100 nm AFM topography images of monolayer MoS$_2$ flakes on SiO$_2$ (a), mica (b) and h-BN (c) substrates, as well as of the bare substrates. Fitting the histograms to a Gaussian distribution (lines) the roughness of the surface can be evaluated by means of the standard deviation (σ). MoS$_2$ single layers deposited on SiO$_2$ substrates show a remarkably higher corrugation than those deposited on mica and, especially on h-BN. The height histogram of a bulk MoS$_2$ surface is shown in the three panels for reference.

The roughness of single layer $MoS_2$ on h-BN follows a remarkably different behavior. We find that the roughness of a single layer $MoS_2$ crystal ($\sigma = 63$ pm) is between that of the bare h-BN substrate ($\sigma = 57$ pm) and that of bulk $MoS_2$ ($\sigma = 66$ pm). We find, therefore, a behavior similar to that reported by C. Dean et al. [18] for graphene on h-BN substrates. However, in the case of graphene, because its bending rigidity (1.4 eV) [25] is lower than for $MoS_2$ (9.6 eV) [24], the roughness of the graphene flakes is indistinguishable from that of the bare h-BN substrate. Interestingly, while pristine mica and h-BN crystals are atomically flat, single-layer $MoS_2$ is rougher when deposited on mica than on h-BN if the transfer is performed under ambient conditions. We attribute this behavior to the strong hydrophilic character of mica [16], as discussed above, while the hydrophobic character of h-BN [27] yields atomically flat single-layer $MoS_2$ even when deposited using a simple transfer technique under ambient conditions.

Given the expected strong dependence of sliding friction on surface roughness we also study the role of the substrate on the friction of single layer $MoS_2$, performing lateral friction force experiments with the tip of an AFM, and find a remarkable relationship between friction and roughness. The friction measurements were carried out at a normal load force of 30 nN. Figures 2b, 2d and 2f show lateral friction force AFM images of $MoS_2$ single layers on $SiO_2$, h-BN and mica respectively, and Figures 2g, 2h and 2i show the corresponding lateral force measurements along a single scan line. The friction force is normalized to its value on the bare $SiO_2$ substrate, 260 nN.

In the case of single-layer MoS$_2$ flakes on SiO$_2$, the lateral friction force is about five times higher than that of a bulk MoS$_2$ crystal. Furthermore, the lateral friction force of MoS$_2$ flakes on SiO$_2$ decreases monotonically towards its limiting bulk value as the flake thickness increases [11], reaching the asymptotic limit for a flake thickness above 5 nm, as shown in Figure 4a. A lower friction force is obtained for single layers on mica, as expected for a flatter MoS$_2$ surface, indicating a strong correlation between flake's roughness and friction.

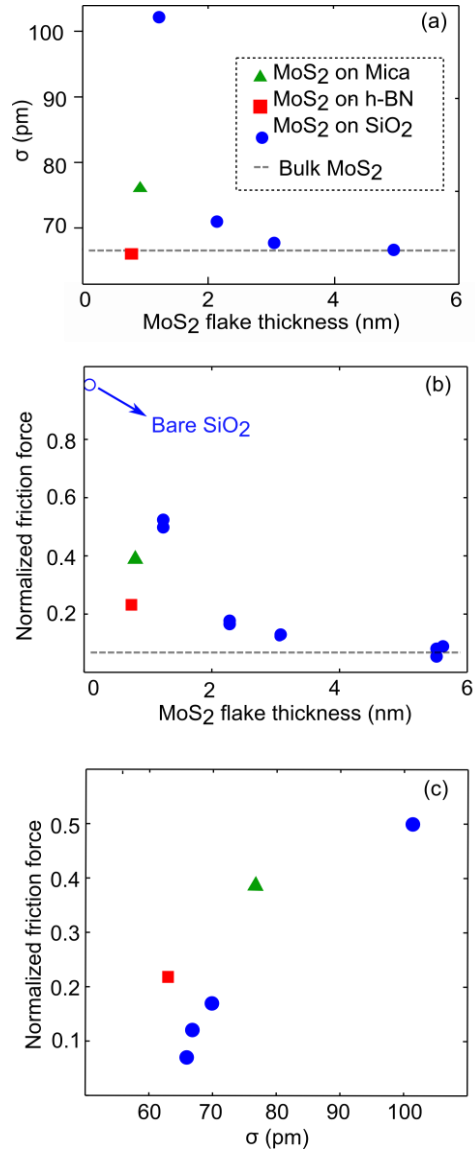

FIG. 4. (a) MoS$_2$ flake thickness dependent surface roughness. The roughness of the MoS$_2$ flake deposited on SiO$_2$ reaches its bulk value for a thickness of 6 nm (9 layers). In the case of MoS$_2$ on h-BN, a roughness comparable with that of bulk MoS$_2$ is obtained even for single layers. (b) AFM lateral friction force for the different stacks of crystals as a function of the MoS$_2$ flake thickness. Again, the friction tends asymptotically to the value for bulk MoS$_2$ as thickness increases. The friction of MoS$_2$ monolayers on mica is lower than on SiO$_2$ and even lower for h-BN substrates. (c) Friction force of MoS$_2$ single layers as a function of the roughness. Notice that for MoS$_2$ flakes on h-BN, even though the roughness is very similar to that of bulk MoS$_2$, the measured friction force is five times lower in the latter case.

However, we find that friction force measurements for MoS$_2$ single layers on h-BN are still noticeable larger than that obtained for the surface of bulk MoS$_2$, even though their roughness is very similar, as shown in Figure 4c. Quantitatively, the roughness ratio is 0.95 while the friction force ratio is 3. A similar surprising behavior has been previously observed by J. Hone *et al.* [26] for graphene crystals on different substrates. It has been attributed to the emergence of a local out-of-plane deformation of the flake due to the tip-flake adhesion force, which leads to an increase of the contact area between the tip and the MoS$_2$ crystal. Thus, for a weak bonding strength between the flake and the substrate the friction force can result significantly increased. Therefore, our results suggest that the bonding strength between the MoS$_2$ flake and the underlying h-BN crystal is lower than between layers of a bulk MoS$_2$ crystal.

In summary, we find that the roughness and sliding friction of MoS$_2$ monolayers can be strongly influenced by the substrate on which they are deposited. MoS$_2$ monolayer flakes have been transferred on top of amorphous SiO$_2$, and on two different atomically flat crystals, mica and hexagonal BN, by a combination of micromechanical cleavage and a deterministic transfer method based on the use of viscoelastic stamps [21]. We have studied the surface roughness and lateral friction force by atomic force microscopy. Compared to MoS$_2$ monolayers on SiO$_2$ substrates, there is a remarkable reduction of the MoS$_2$ flake roughness when deposited on mica and, especially, on h-BN substrates. In the latter case the roughness is five times lower than for MoS$_2$ on SiO$_2$, and comparable to that of the surface of bulk MoS$_2$ crystals. In addition we determine the relationship between lateral friction force and surface roughness of MoS$_2$ either on SiO$_2$, mica or h-BN. We find that the friction force of MoS$_2$ on mica is lower than for MoS$_2$ on SiO$_2$, as expected for the observed lower surface roughness. An even stronger friction force reduction is found for MoS$_2$ on h-BN. However, remarkably, even if the roughness in this case is comparable to that of bulk MoS$_2$, the friction force is at least two times larger. This can be attributed to the emergence of a local out-of-plane deformation of the MoS$_2$/h-BN interface due to the tip-flake adhesion force [26], which leads to an increase of the contact area between the tip and the MoS$_2$ crystal. In conclusion, h-BN thin crystals are promising and advantageous substrates for single layer MoS$_2$ devices because of the remarkable reduction of single-layer MoS$_2$ surface roughness, sliding friction and local inhomogeneities. The quantitative analysis of surface roughness and sliding friction can be used as a powerful method to characterize the quality of ultra-thin van der Waals heterostructures.

**ACKNOWLEDGMENTS**

J.Q., N.A. and G.R-B. gratefully acknowledge financial support by MICINN/MINECO (Spain) through program MAT2011-25046 and Comunidad de Madrid through program Nanobiomagnet S2009/MAT-1726. A.C-G. acknowledges financial support of the European Union Seventh Framework program through the FP7-Marie Curie Project PIEF-GA-2011-300802 ('STRENGTHNANO').